\RequirePackage[displaymath]{lineno}
\documentclass[twocolumn,aps,prc,showpacs,superscriptaddress,floatfix,nofootinbib]{revtex4-2}
\usepackage{newtxtext,newtxmath,booktabs,siunitx}
\usepackage{dcolumn}% Align table columns on decimal point
\usepackage{bm}% bold math
\usepackage{float}
\usepackage{ulem}
\usepackage[]{graphicx}  % remove 'demo' option for your real document
\usepackage{booktabs}
\usepackage{tabularx}
\usepackage{amsmath}
\usepackage{xcolor}
\usepackage{multirow}
\usepackage[colorlinks,citecolor=blue,urlcolor=blue,linkcolor=blue]{hyperref}
\usepackage{booktabs}

\newcommand{\snn}{\sqrt{s_{\rm NN}}}

\newcommand{\OO}{$^{16}$O +$^{16}$O}
\newcommand{\Oxy}{$^{16}$O}
\newcommand {\Nch}      {N_{\rm ch}}

\newcommand{\trento}{T\raisebox{-0.5ex}{R}ENTo}

\newcommand{\iebe} {{\tt iEBE-VISHNU}}

\newcommand{\vvt}{v_{2}\{2\}/v_{3}\{2\}} 
\newcommand{\vvf}{v_{2}\{2\}/v_{2}\{4\}} 
\newcommand{\eet}{\epsilon_{2}\{2\}/\epsilon_{3}\{2\}}
\newcommand{\eef}{\epsilon_{2}\{2\}/\epsilon_{2}\{4\}}

\begin{document}

\title{Probing the tetrahedral $\alpha$ clusters in relativistic $^{16}$O + $^{16}$O collisions}

\author{Jin-Yu Hu}
\affiliation{Department of Modern Physics, University of Science and Technology of China, Anhui 230026, China}
\affiliation{School of Science, Huzhou Normal University, Huzhou, Zhejiang 313000, China}

\author{Hao-jie Xu}
\email{Corresponding author: haojiexu@zjhu.edu.cn}
\affiliation{School of Science, Huzhou Normal University, Huzhou, Zhejiang 313000, China}
\affiliation{Strong-Coupling Physics International Research Laboratory (SPiRL), Huzhou Normal University, Huzhou, Zhejiang 313000, China}

\author{Xiaobao Wang}
\email{Corresponding author: xbwang@zjhu.edu.cn}
\affiliation{School of Science, Huzhou Normal University, Huzhou, Zhejiang 313000, China}

\author{Shi Pu}
\email{Corresponding author: shipu@ustc.edu.cn}
\affiliation{Department of Modern Physics, University of Science and Technology of China, Anhui 230026, China}
\affiliation{Southern Center for Nuclear-Science Theory (SCNT), Institute of Modern Physics, Chinese Academy of Sciences, Huizhou 516000, Guangdong Province, China}

\begin{abstract}
Relativistic \OO\ collisions provide a valuable opportunity to study both the quark-gluon plasma formed in small systems and the intrinsic structure of \Oxy. Recent implementations of \textit{ab initio} nuclear configurations in heavy-ion simulations have produced different predictions for cluster-sensitive observables, making it difficult to identify the origin of possible $\alpha$-cluster signals. In this work, we introduce a controlled sampling scheme that varies the compactness of $\alpha$-cluster-induced multi-nucleon correlations while keeping the one-body density distribution of \Oxy\ fixed. This allows us to separate effects driven by the tetrahedral one-body density from those driven by genuine multi-nucleon correlations. We show that the normalized ratios ${\rm Norm}(v_{2}\{2\}/v_{2}\{4\})$ and ${\rm Norm}(v_{2}\{2\}/v_{3}\{2\})$, together with their initial-state eccentricity counterparts, provide complementary constraints on initial-condition model dependence and cluster compactness. Hybrid hydrodynamic simulations and comparisons with recent LHC measurements further clarify the extent to which the initial-state signals survive final-state evolution. Our results provide a framework for using relativistic light-ion collisions to constrain cluster correlations in \Oxy.
\end{abstract}

\maketitle

\section{Introduction}
Relativistic heavy ion collisions produce an extremely hot and dense state of quantum chromodynamics (QCD) matter known as the Quark Gluon Plasma (QGP)~\cite{Adcox:2004mh,Adams:2005dq,ALICE:2010suc,Shuryak:2008eq}. 
The observation of strong anisotropic flow in the momentum distributions of final-state hadrons provides compelling evidence that the QGP behaves as a nearly perfect fluid — a phenomenon that is well-described by relativistic hydrodynamic models~\cite{Ollitrault:1992bk,Kovtun:2004de,Romatschke:2007mq,Shen:2014vra,Song:2017wtw}. 
Recently, relativistic oxygen-oxygen (\OO) collisions conducted at both the Relativistic Heavy Ion Collider (RHIC)~\cite{Huang:2023viw} and the Large Hadron Collider (LHC)~\cite{ALICE-Oxygen} provide new data to deepen our understanding of QGP evolution in small systems~\cite{Zhao:2017rgg,Noronha:2024dtq}.
To investigate the evolution of the QGP, precise knowledge of the initial conditions is essential.

For light nuclei such as \Oxy, the existence of $\alpha$ clusters, a key aspect of nuclear structure, remains debated \cite{Wheeler:1937zza, Dennison:1954zz, FESHBACH19737, Robson:1979zz, Bauhoff:1984zza, Tohsaki:2001an, Yakovlev:1997vi, Bijker_AIP,Epelbaum:2013paa}.
Several decades ago, the tetrahedral arrangement of four interacting $\alpha$ clusters was proposed as the ground state (g.s.) structure of $^{16}$O \cite{Robson:1979zz, Pauling_comment_PRL}.
Recent studies, as referenced in \cite{Iachello_PRL_2014}%[Iachello]
, have further supported this hypothesis. This is primarily based on observations of a tetrahedral rotor characterized by its distinctive level sequence: the lowest 0$^+$, followed by the lowest 3$^-$, 4$^+$, and 6$^+$ states. Multiple experimental indicators, including energy levels, charge form factors, elastic scattering data, and $EL$ transition data, collectively suggest the existence of this tetrahedral symmetry \cite{Robson:1979zz,Pauling_comment_PRL,Iachello_PRL_2014}.
Furthermore, energy density functional theory (DFT) calculations incorporating symmetry restoration have independently confirmed that the tetrahedral configuration represents the g.s. of \Oxy~\cite{WANG2019498}.

A recent breakthrough has revealed that the structure of colliding nuclei can be extracted in relativistic heavy ion collision with instantaneous snapshots of the collision geometry~\cite{Li:2019kkh,Xu:2021uar,Zhang:2021kxj,Ryssens:2023fkv,Xu:2024bdh,STAR:2024wgy,Giacalone:2024ixe}.
Several \textit{ab initio} calculations, such as Variational Monte Carlo (VMC), Nuclear Lattice Effective Field Theory (NLEFT), and Extended Quantum Molecule Dynamics (EQMD), supporting the existence of $\alpha$ clusters in nuclei~\cite{Epelbaum:2013paa,Pieper:2002ne,Lee:2008fa}, have been applied to relativistic heavy-ion simulations~\cite{Rybczynski:2019adt,Behera:2021zhi,Ding:2023ibq,Liu:2023gun,Zhang:2024vkh,Mehrabpour:2025ogw,Brewer:2021kiv,Summerfield:2021oex}.
The timing data available in heavy-ion collisions has prompted considerable efforts to identify signals of these clusters in relativistic \OO\ collisions using configurations derived from these theoretical frameworks.
However, divergent predictions for observables from different configurations complicate the interpretation of potential cluster effects~\cite{Huang:2023viw}. As indicated in Ref.~\cite{YuanyuanWang:2024sgp}, these discrepancies arise from variations in the compactness of $\alpha$ clusters in oxygen with tetrahedral configurations.
Such a modification of compactness also alters the one-body density distribution accordingly.

Unlike heavy nuclei, light nuclei generally exhibit lower density and possess weaker binding energy. Consequently, they can gain energy through the formation of strongly bound clusters, e.g. $\alpha$~\cite{Wheeler:1937zza,VONOERTZEN200643,Ebran:2012ww,Schuck2013}. This characteristic makes many-body correlations particularly significant in light nuclei systems. Therefore, a description based on the one-body density as the only degree of freedom may prove inadequate for light nuclei, necessitating the inclusion of multi-nucleon correlations.
Specifically, the tetrahedral configuration of $\alpha$ clusters for \Oxy\ contributes to initial conditions in two distinct ways: through the $Y_{32}$ octupole deformation of the one-body density distribution and via multi-nucleon correlations from cluster correlations.
To fully understand how cluster configurations influence %affect 
final observables in relativistic \OO\ collisions, 
multi-nucleon correlation effects must be distinguished from contributions arising solely from the one-body density.
For example, cluster signatures can emerge through event-by-event nucleon-nucleon interactions even in configurations where the one-body density appears spherical. Furthermore, model dependence requires careful
investigation, as different initial geometry models yield significantly varied predictions despite using identical cluster configurations for oxygen~\cite{Zhang:2024vkh}.

With recent relativistic \OO\ collision data~\cite{STAR:2025ivi,CMS:2025tga,ALICE:2025luc,ATLAS:2025nnt}, it is imperative to address these obstacles within heavy-ion models. In this work, we employ ground-state $^{16}$O density profiles from Skyrme-DFT calculations, contrasting configurations with and without tetrahedral symmetry. One configuration features a pure $\hat{Q}_{32} = 40$ fm$^3$ moment (octupole deformation $\beta_{32} = 0.339$ with a $Y_{32}$ shape, which is the g.s. of $^{16}$O given by the symmetry-restored potential energy surface (PES), as explained in Ref.~\cite{WANG2019498}) and a spherical configuration (the g.s. of $^{16}$O given by PES without symmetry restoration). 
This study employs heavy-ion models to explicitly investigate their sensitivity to initial-state assumptions, providing timely insights into cluster effects in relativistic \OO\ collisions.

\bigskip

%{\em {\it Model setups.}}
\section{Model setups}
In this study, we investigate the second- and third-order flow harmonics in central relativistic \OO\ collisions at $\snn=5.36$~TeV. The inelastic nucleon-nucleon cross section is set to $\sigma_{\rm NN} = 70$~mb. Leveraging the well-established linear response in hydrodynamic simulations $v_{n}=\kappa_n\epsilon_{n}$ ($n=2,3$)~\cite{Qiu:2011iv}, we employ initial geometry models that include Monte Carlo (MC) Glauber~\cite{Loizides:2014vua} and \trento\ \cite{Moreland:2014oya}. Here $v_{n}$ is the $n$-th order flow harmonics and $\epsilon_{n}$ denotes the corresponding initial spatial eccentricity~\cite{Xu:2021vpn,Wang:2023yis}. 
Centrality is determined by initial entropy density (or the so-called multiplicity $N_{\rm ch}$), where $\Nch \propto (1-x)N_{\rm part}/2 + xN_{\rm coll}$ with $x=0.1$ in the MC Glauber model and $\Nch = \int dx\,dy\,\left[(T_A^p + T_B^p)/2\right]^{1/p}$ with $p=0$ in \trento\ model.
Here, $N_{\rm part}$ and $N_{\rm coll}$ denote participant nucleons and binary collisions, while $T_A$, $T_B$ are nuclear thickness functions. We validate the linear hydrodynamic response using (2 + 1) dimensional viscous hydrodynamics via \iebe\ simulations \cite{Song:2010aq,Shen:2014vra}.
The model parameters for both the initial state (\trento) and the hydrodynamic evolution are listed in Table~\ref{tab:hydro_paras}. 
They are guided by the Bayesian-calibrated setup of Ref.~\cite{Moreland:2018gsh} and are adjusted in this work to provide a reasonable description of the recent \OO\ data~\cite{CMS:2025tga}.
We note that the \trento\ parameters are applied to the constituent \trento\ with constituent number $m=6$, and we have checked that the difference between the nucleon \trento ($m=1$) and the constituent \trento ($m=6$) models is negligible.

\begin{figure}[t]
    \includegraphics[width=0.48\textwidth]{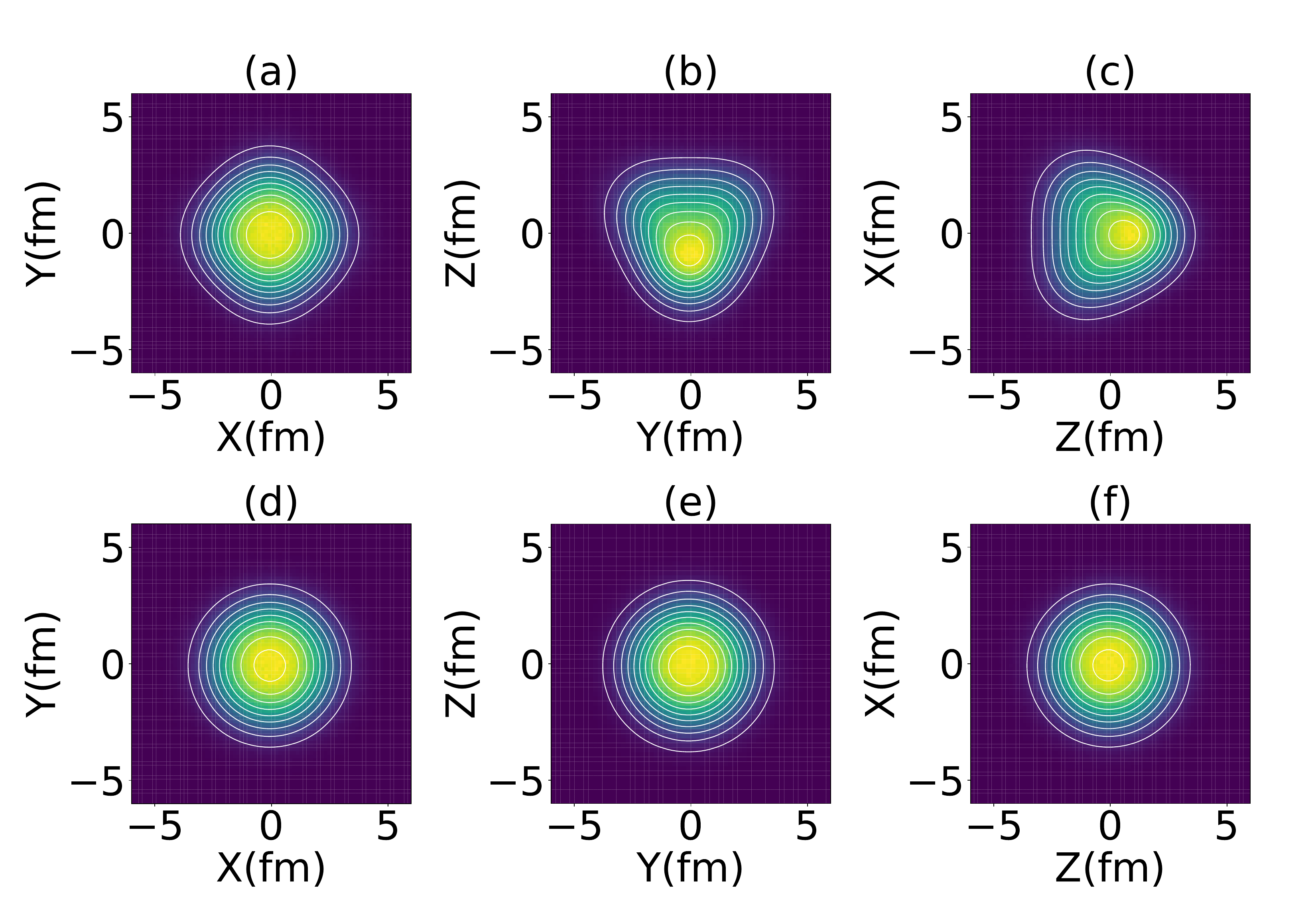}
    \caption{(Color online) 
   One-body density profiles of \Oxy\ from Skyrme energy density functional calculations with ($\hat{Q}_{32}=40$ fm$^3$, upper panels a-c) and without ($\hat{Q}_{32}=0$, lower panels d-f) tetrahedral symmetry.
  }
\label{fig:onebodydensity}
\end{figure}

\begin{table}[h]
\centering
\caption{Input parameters of the initial-state and hydrodynamic models used in this work. 
	The parameter choices are guided by Ref.~\cite{Moreland:2018gsh} and are adjusted to provide a reasonable description of the recent \OO\ data~\cite{CMS:2025tga}.}
\label{tab:hydro_paras}
%\begin{tabular}{p{0.18\columnwidth} p{0.25\columnwidth} p{0.25\columnwidth} p{0.18\columnwidth}}
\begin{tabular}{llll}
\toprule 
\multicolumn{2}{l}{Initial condition/Preeq.} & \multicolumn{2}{l}{QGP medium} \\
\midrule
Norm                    & 20.0               &$(\eta/s)_\mathrm{hrg}$       & $0.2$ \\
$p$                     & 0.0                &$(\eta/s)_\mathrm{min}$       & $0.2$ \\
$k$                     & 0.19               &$(\eta/s)_\mathrm{slope}$     & $1.6\ \mathrm{GeV}^{-1}$ \\
$r_\mathrm{cp}$         & 0.5 fm             &$(\eta/s)_\mathrm{crv}$       & $-0.29$ \\
$n_c$                   & 6.0                &$(\zeta/s)_\mathrm{max}$      & $0.032$ \\
$\omega_{\mathrm{c}}$   &0.43 fm             & $(\zeta/s)_\mathrm{width}$   & $0.024\ \mathrm{GeV}$ \\
$\tau_{\mathrm{fs}}$    &$0.37\mathrm{fm}/c$ & $(\zeta/s)_{T_\mathrm{peak}}$& $0.175\ \mathrm{GeV}$ \\
                        &                    &$T_{\mathrm{switch}}$         & $0.151\ \mathrm{GeV}$ \\
\bottomrule   
\end{tabular}
\end{table}

The density profiles in g.s. of \Oxy\ are obtained from the Skyrme energy density functional calculation. For details, we refer to Ref.~\cite{WANG2019498}.
The density profiles of \Oxy\ differ significantly with and without tetrahedral symmetry, as shown in Fig.~\ref{fig:onebodydensity}. When sampling nucleons based solely on densities from DFT calculations, we capture only mean-field-level contributions to nucleon-nucleon correlations. To investigate cluster effects, substantial beyond-mean-field correlations must be explicitly included.
In this work, we introduce an effective compactness parameter $\chi$ to quantify $\alpha$ cluster correlations. Suppose that the single nucleon distribution function is $f\left(\mathbf{r}\right)$,
we sample four nucleons in each of the four clusters using the weighted distribution function $f_{i}\left(\mathbf{r}\right)=\omega_{i}\left(\mathbf{r}\right)f\left(\mathbf{r}\right)$ $\left(i=1,2,3,4\right)$, where 
\begin{equation}
%\omega_{i}\left(\mathbf{r}\right)=\frac{\exp\left(-\beta\left\Vert \mathbf{r}-\mathbf{C}_{i}\right\Vert ^{2}\right)}{\sum_{j=1}^{4}\exp\left(-\beta\left\Vert \mathbf{r}-\mathbf{C}_{j}\right\Vert ^{2}\right)},
\omega_{i}\left(\mathbf{r}\right)=\frac{\exp\left[-\chi(\mathbf{r}-\mathbf{C}_{i})^{2}\right]}{\sum_{j=1}^{4}\exp\left[-\chi(\mathbf{r}-\mathbf{C}_{j})^{2}\right]},
\end{equation}
and $\mathbf{C}_{i}$ are the centers of the four clusters (for the spherical case, we also assume that it has the same centers as in the tetrahedron case). Since $\sum_{i}\omega_{i}\left(\mathbf{r}\right) = 1$, different $\chi$ will yield the same one-body distribution. The parameter $\chi$ can control
the compactness of the cluster. 
We emphasize that $\chi=0$ corresponds to the one-body distribution without corrections in sampling nucleons, while $\chi \neq 0$ means including the multi-nucleon corrections.
When $\chi$ varies from $0$ to $2$, 
the root mean square (RMS) of a cluster
in the oxygen nucleus obtained from sampling ranges from $2.37\; \mathrm{fm}$
to $1.60 \; \mathrm{fm}$ for $\hat{Q}_{32}=40 \; \mathrm{fm}^{3}$ case (with tetrahedral symmetry)
and from $2.25 \; \mathrm{fm}$ to $1.62 \; \mathrm{fm}$ in the spherical case . 
Increasing $\chi$ further will not alter the size of the clusters in nuclei. Therefore, under an appropriate fixed one-body density, the very small cluster size found in Ref.~\cite{YuanyuanWang:2024sgp} is unlikely. We also verified that the observables will not change with a further increase in the value of $\chi$.

Since we are dealing exclusively with \OO\ collisions, where individual observables are sensitive to both the initial geometry and the medium response, we focus on the ratios $\eet$ and $\eef$ and on their final-state counterparts $\vvt$ and $\vvf$. The eccentricity cumulants are defined in analogy with the standard flow cumulants, for example
\begin{equation}
\epsilon_n\{2\}=\sqrt{\langle \epsilon_n^2\rangle},\qquad
\epsilon_n\{4\}=\left(2\langle \epsilon_n^2\rangle^2-\langle \epsilon_n^4\rangle\right)^{1/4},
\end{equation}
with $v_n\{2\}$ and $v_n\{4\}$ defined by the corresponding two- and four-particle azimuthal cumulants~\cite{Bilandzic:2010jr,Bilandzic:2013kga}. 
Similar flow-cumulant ratios have been proposed as probes of $\alpha$ clustering in relativistic \OO\ collisions~\cite{Summerfield:2021oex,YuanyuanWang:2024sgp}.
To reduce common systematic uncertainties and emphasize centrality-dependent trends, we introduce 
\begin{equation}
{\rm Norm}(X) = \frac{X_{[\rm given\ centrality]}}{X_{[0-1\%\ \rm centrality]}} .
\end{equation}
The same normalization is applied to both eccentricity and flow ratios. We note that relativistic p+$^{16}$O collisions could provide additional constraints on the initial condition and reduce systematic uncertainties; we leave such an investigation for future work.

\begin{figure*}[hbt]
    \includegraphics[width=0.8\textwidth]{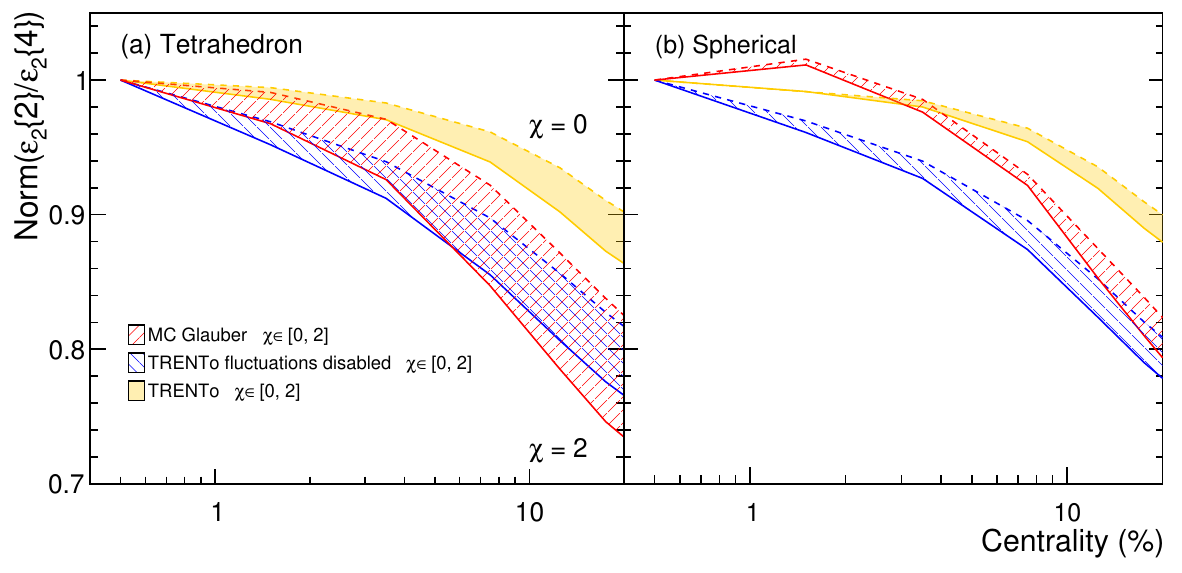}
    \caption{(Color online) 
  Normalized $\eef$ ratios for \OO\ collisions at $\snn=5.36$~TeV. 
Predictions are shown for both the MC Glauber model and \trento\ model 
with varying effective compactness parameter $\chi$. 
\trento\ simulations with gamma fluctuations disabled are also presented.
  }
\label{fig:e24e22}
\end{figure*}

\bigskip

%{\em{\it Results of initial state.}}
\section{Results of initial state}
The normalized ratios $\eef$ for \OO\ collisions at $\snn=5.36$~ TeV, calculated using initial geometry models, are presented in Fig.~\ref{fig:e24e22}.  In Ref.~\cite{Huang:2023viw},  the centrality dependence of $\eef$ from MC Glauber simulations was compared to STAR preliminary data at $\snn=200$~GeV. Such centrality-dependent trends have been posited as possible indicators of cluster effects, since the predictions differ from various \textit{ab initio} density distributions. Our MC Glauber results at $\snn=5.36$~TeV (red bands) agree with RHIC energy predictions despite employing distinct density-calculation methods. However, the differences between the spherical and clustered configurations are overwhelmed by the model dependence compared to \trento\ simulations (yellow bands) using identical configurations.

This substantial dependence on the model has been determined to be primarily attributable to the fluctuation parameter $k$ in the source weighting scheme of the default \trento. In the \trento\ simulations, a weight factor for entropy deposition is assigned to each participant~\cite{Bozek:2013uha,Moreland:2014oya}. This factor is generated from a $\Gamma$ distribution, with a mean of $1$ and a variance of $1/k$. These fluctuations are implemented to reproduce multiplicity distributions in small collision systems. Previous research using the default \trento\ model indicates that the value of $k$ is generally set at approximately $1$. We find that, when these weight fluctuations are disabled ($k\rightarrow\infty$) for relativistic \OO\ collisions, \trento\ simulations (illustrated by the blue bands in Fig.~\ref{fig:e24e22}) yield results consistent with the MC Glauber predictions. These different scenarios of the weight factor can be readily distinguished by experimental ${\rm Norm}(\vvf)$ data.

Having clarified the model dependence, we now focus on the cluster effects in \OO\ collisions. In all model simulations, the centrality dependence of observables becomes more pronounced in the top $15\%$ centrality range as the effective compactness parameter $\chi$ increases. This aligns with previous investigations of $\alpha$ cluster compactness~\cite{YuanyuanWang:2024sgp}, though our study maintains rigorous constraints by preserving identical one-body density distributions across configurations. We find that one-body density distributions have a negligible impact on ${\rm Norm}(\eef)$ when multi-nucleon correlations are absent
($\chi=0$), where nucleons are sampled
independently from the density profile. In contrast, multi-nucleon correlations ($\chi=2$) exhibit significant amplification for tetrahedrally symmetric densities. 
However, this correlation-driven effect is obscured by model dependence (see the red and blue bands in Fig.~\ref{fig:e24e22}) in relativistic \OO\ collisions. 
We note that nucleon size is another source of uncertainty, as $\epsilon_{2}\{2\}$ and $\epsilon_{2}\{4\}$ demonstrate opposite sensitivities to initial state fluctuations.

\begin{figure*}[hbt]
    \includegraphics[width=0.8\textwidth]{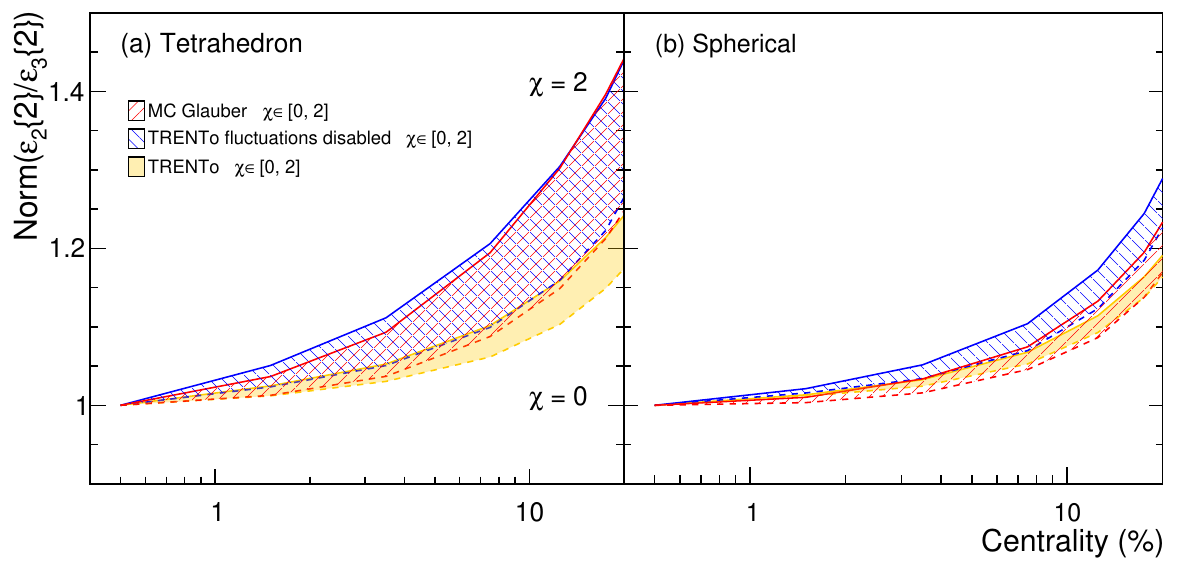}
    \caption{(Color online) 
    Similar to Fig.~\ref{fig:e24e22} but for the normalized $\eet$ ratios.
  }
\label{fig:e32e22}
\end{figure*}

Tetrahedral symmetry implies a non-vanishing octupole deformation that contributes significantly to third-order flow harmonics ($v_3$) in most central collisions. We present normalized $\eet$ ratios in Fig.~\ref{fig:e32e22}. For ${\rm Norm}(\eet)$, one-body distributions with and without tetrahedral symmetry exhibit distinct behaviors: predictions for centrality-dependent trends are significantly sharper with tetrahedral deformation than with spherical symmetry. 
Unfortunately, we notice that these differences can be compensated for by multi-nucleon correlations in the spherical case. For example, the spherical configuration with $\chi = 2$ roughly reproduces the trend in the deformed case with $\chi = 0$.
Consequently, a sharp trend observed experimentally cannot be attributed to density profiles with small $\hat{Q}_{32}$ and weak multi-nucleon correlations.
However, if fluctuating weighting factors dominate (yellow bands), exploring cluster effects through ${\rm Norm}(\vvt)$ observables in relativistic \OO\ collisions becomes challenging. 
Thus, refining the heavy-ion model is necessary to explore the cluster effects in relativistic heavy-ion collisions.

\begin{figure}[hbt]
    \includegraphics[width=0.4\textwidth]{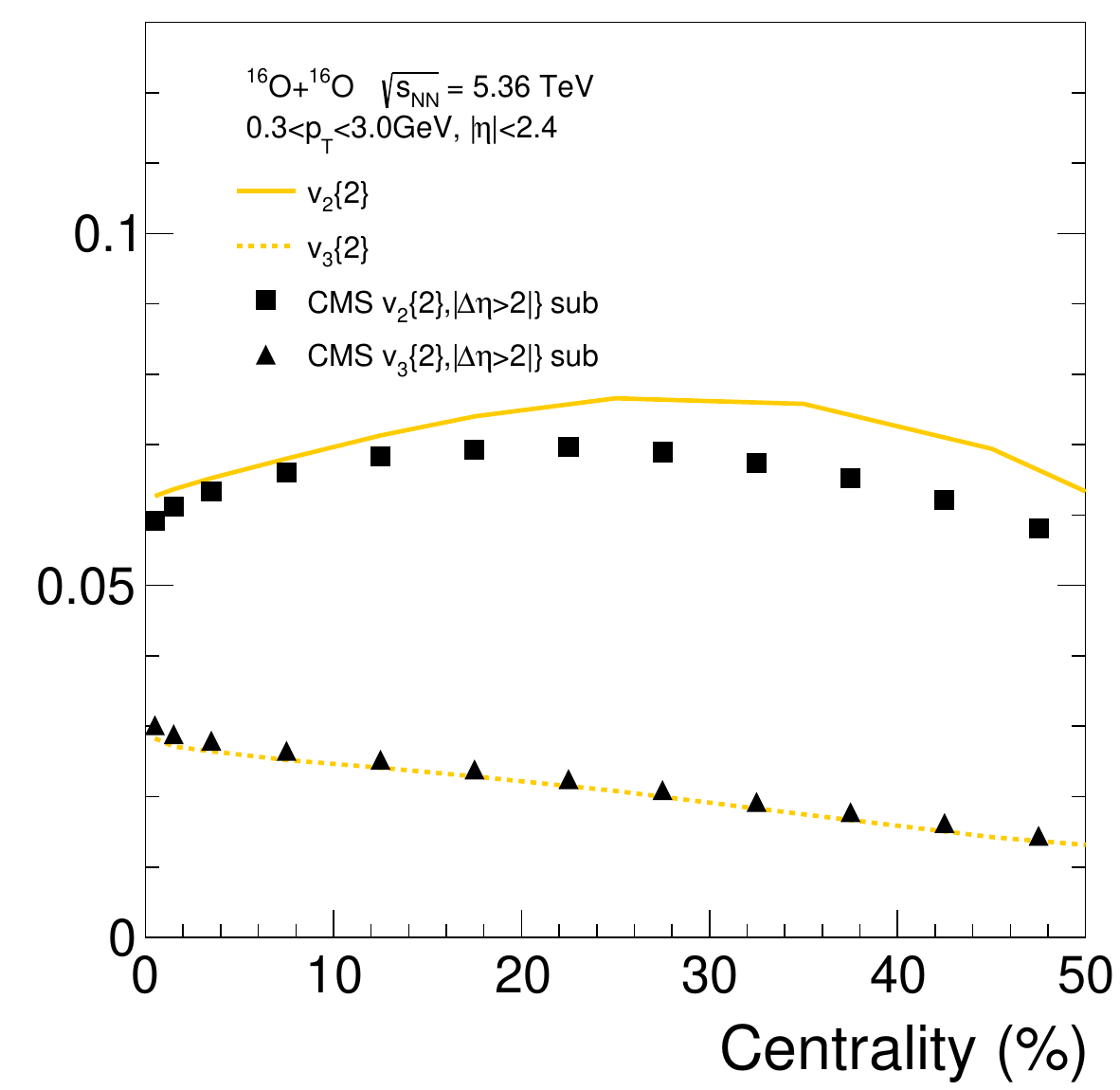}
    \caption{The comparison of hybrid model simulation with the experimental results of $v_2\{2\}$ and $v_3\{2\}$ from~\cite{CMS:2025tga}. 
  }
\label{fig:comparison with exp}
\end{figure}

\begin{figure}[hbt]
    \includegraphics[width=0.4\textwidth]{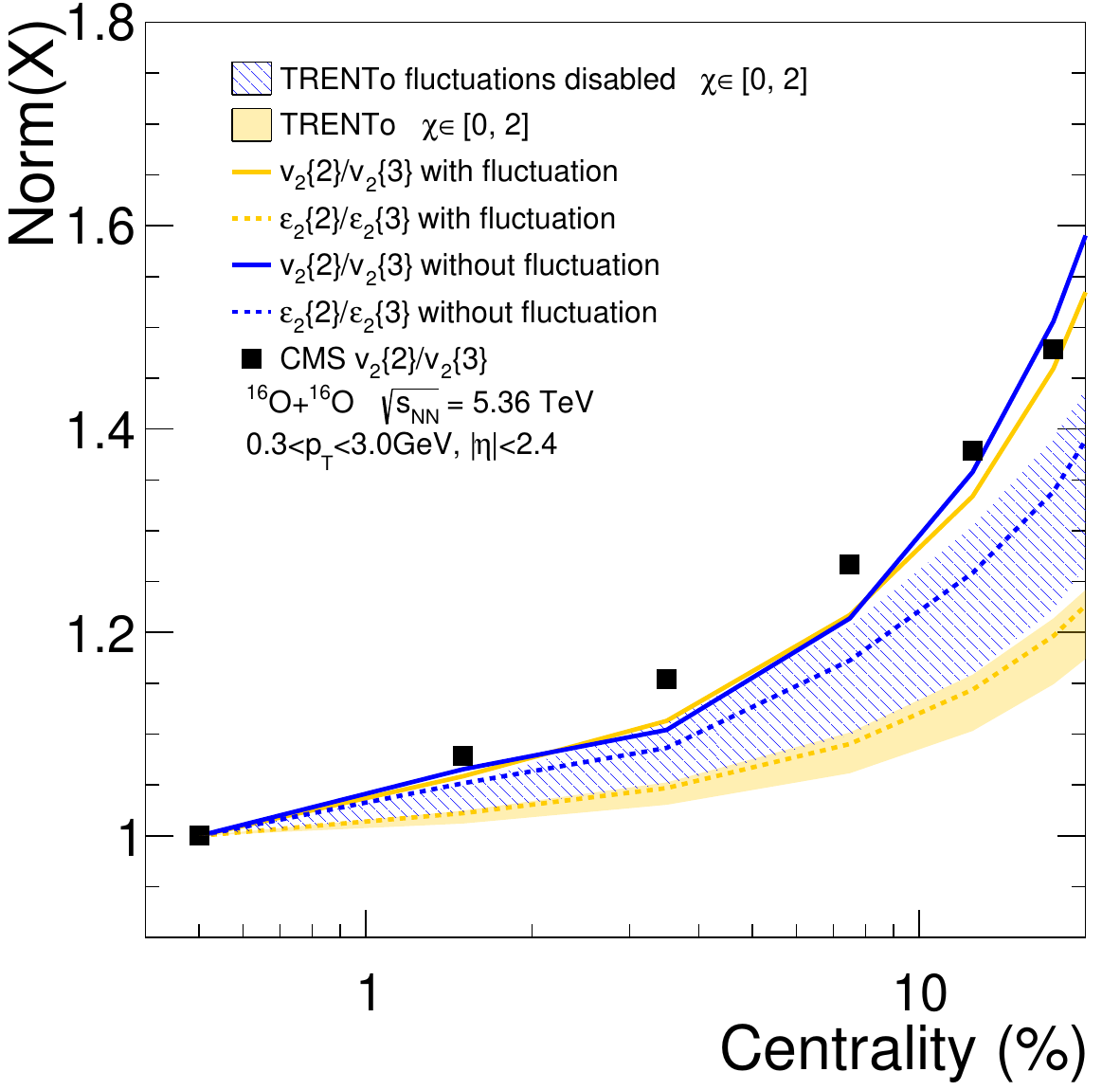}
    \caption{Predictions with VMC densities for the normalized $v_2\{2\}/v_3\{2\}$ ratio in relativistic \OO\ collisions. Dashed and solid curves denote the initial-state eccentricity predictor and the final-state hybrid hydrodynamic results, respectively. Different colors indicate whether the \trento\ weight fluctuations are included. The black points show the corresponding recent LHC measurement~\cite{CMS:2025tga}. 
  }
\label{fig:VMC}
\end{figure}

Although we employ DFT-derived density profiles as the main controlled setup, our aim is not to determine which microscopic nuclear calculation gives the best description of \Oxy. Rather, the DFT densities provide two limiting cases with and without tetrahedral one-body deformation. The value $\hat{Q}_{32}=40~\mathrm{fm}^3$ is taken from the symmetry-restored Skyrme-DFT calculation for the ground state of \Oxy; a larger $\hat{Q}_{32}$ would lead to a stronger centrality dependence in both ${\rm Norm}(\vvt)$ and ${\rm Norm}(\vvf)$.

To connect the controlled study to microscopic configurations, we also implement VMC density profiles in \trento\ simulations. The details of the VMC calculations can be found in Ref.~\cite{RevModPhys.87.1067}. The resulting centrality dependence of ${\rm Norm}(\eet)$ is close to the trend generated by configurations with pronounced $Y_{32}$ deformation and strong multi-nucleon correlations, i.e., large $\chi$ and $\hat{Q}_{32}$, and it differs from the spherical reference calculation. This comparison illustrates how the proposed observables can be used to constrain the compactness of cluster correlations once the initial-condition model dependence is under control.

\bigskip

\section{Results of final state}
So far, our analysis has focused on initial-state eccentricities. The use of eccentricity ratios as predictors of flow ratios relies on an approximate hydrodynamic response~\cite{Noronha-Hostler:2015dbi}, which must be checked explicitly in a small system such as \OO\ collisions. We therefore perform hybrid simulations with the \iebe\ model for $5\times 10^6$ hydrodynamic events of \OO\ collisions at $\snn=5.36$~TeV using VMC densities, followed by ten oversamplings of the {\tt UrQMD} afterburner for each hydrodynamic event.

\begin{figure}
    \centering
    \includegraphics[width=0.8\linewidth]{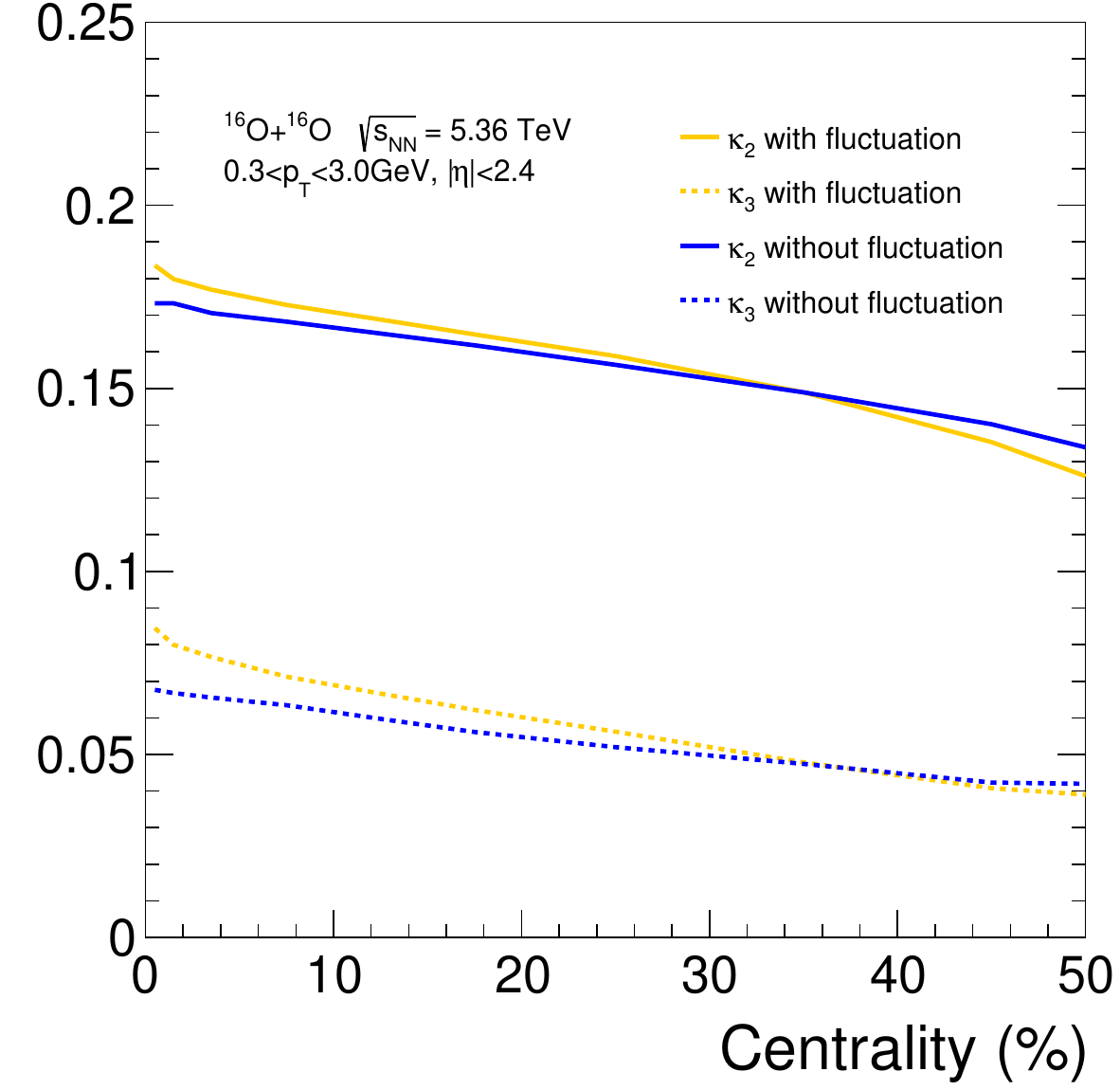}
    \caption{Hydrodynamic response of the second- and third-order flow harmonics in \OO\ collisions. The comparison tests the applicability of the linear-response picture in this small collision system.}
    \label{fig:response}
\end{figure}

\begin{figure}
    \centering
    \includegraphics[width=0.8\linewidth]{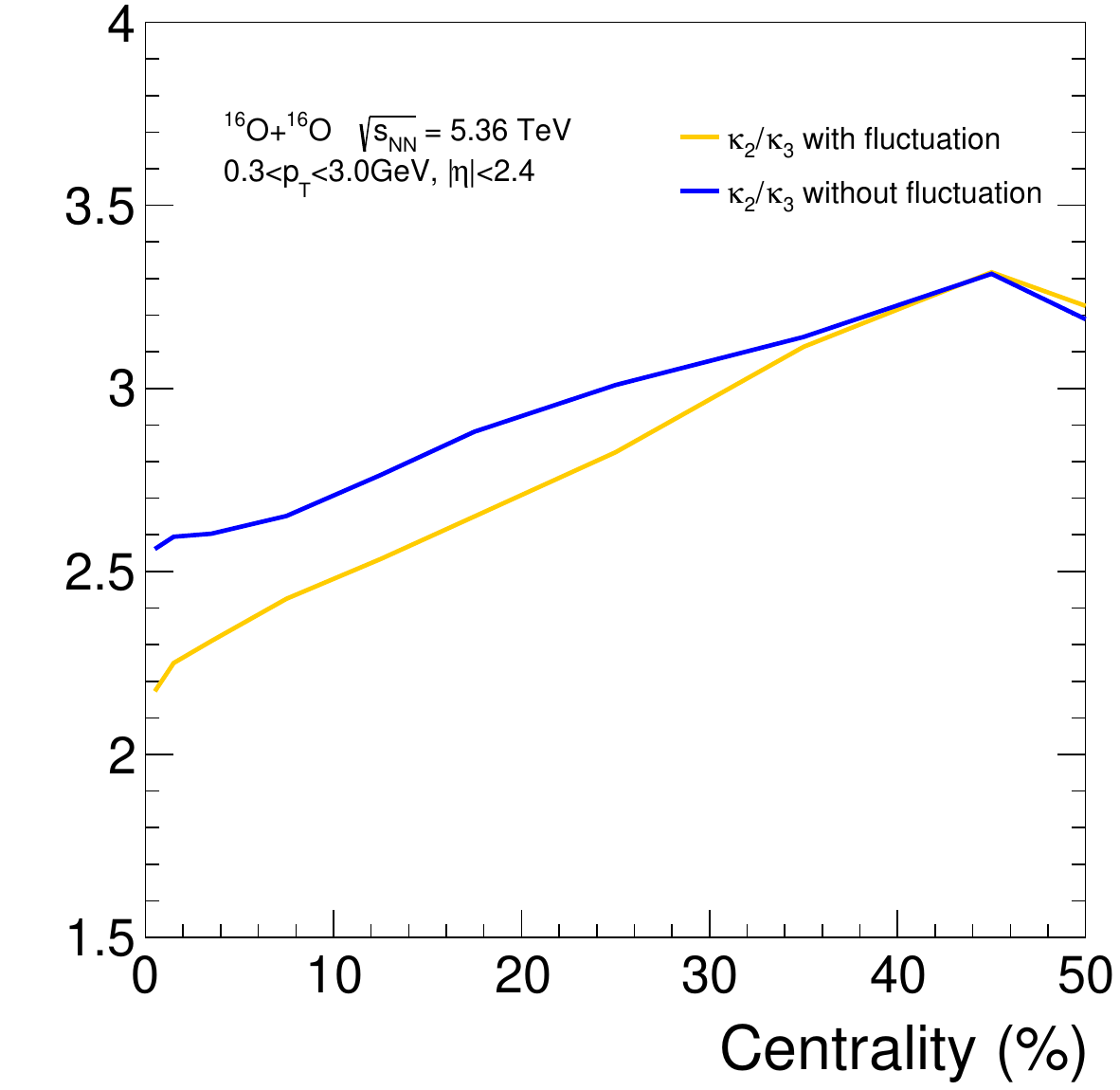}
    \caption{Centrality dependence of the response ratio $\kappa_{2}/\kappa_{3}$. Its variation explains why the normalized final-state $v_2\{2\}/v_3\{2\}$ ratio does not exactly follow the corresponding initial-state eccentricity ratio.}
    \label{fig:responseratio}
\end{figure}

Because the statistical uncertainty of $v_2\{4\}$ remains large with the present event statistics, we focus on $v_2\{2\}$, $v_3\{2\}$, and their normalized ratio $\vvt$. The absolute values of $v_2\{2\}$ and $v_3\{2\}$ are compared with recent LHC measurements in Fig.~\ref{fig:comparison with exp}. The calculation gives a reasonable description of the measured magnitudes within the present model uncertainty. Figure~\ref{fig:VMC} compares the normalized $\vvt$ ratio from the initial-state predictor with the result after hybrid hydrodynamic evolution. The two are not identical, because the response coefficients for $v_2$ and $v_3$ have different centrality dependences, see Fig.~\ref{fig:response}. Equivalently, the ratio $\kappa_2/\kappa_3$ is not constant, as shown in Fig.~\ref{fig:responseratio}.

We also find that the final-state normalized $\vvt$ ratio is less sensitive to the \trento\ weight fluctuations than the corresponding initial-state eccentricity ratio. This reflects a nontrivial interplay between initial fluctuations and final-state response in small systems, different from the behavior usually expected in large systems such as Pb+Pb collisions. These results demonstrate that eccentricity ratios remain useful diagnostic tools, but a quantitative extraction of cluster structure from data requires simultaneous control of the initial condition and the hydrodynamic response.

\bigskip

\section{Summary}
We have investigated how tetrahedral one-body deformation and $\alpha$-cluster-induced multi-nucleon correlations in \Oxy\ can be probed in relativistic \OO\ collisions. The central idea of this work is to vary the compactness of the clusters through the parameter $\chi$ while keeping the one-body density fixed. This construction separates effects caused by the average tetrahedral density from those caused by genuine multi-nucleon correlations.

Our initial-state analysis shows that ${\rm Norm}(\eef)$ is only weakly affected by the tetrahedral octupole deformation itself, but is sensitive to initial-condition model ingredients, especially the $\Gamma$ weight fluctuations in \trento. When these weight fluctuations are disabled, the \trento\ results move toward the MC Glauber results. In contrast, ${\rm Norm}(\eet)$ is more directly sensitive to the tetrahedral $Y_{32}$ geometry and to compact $\alpha$-cluster correlations, although the two effects can partly compensate each other. This compensation demonstrates the necessity of a controlled framework in which one-body density and multi-nucleon correlations can be varied independently.

Hybrid hydrodynamic simulations with VMC densities show that the absolute values of $v_2\{2\}$ and $v_3\{2\}$ are compatible with recent LHC measurements within current model uncertainties. They also show that the mapping from eccentricity ratios to flow ratios is not exact in small systems because $\kappa_2/\kappa_3$ depends on centrality. Therefore, the proposed normalized ratios should be used together with hydrodynamic response studies and experimental data. Our results provide a practical framework for constraining the compactness of $\alpha$ clusters and the tetrahedral structure of \Oxy\ with relativistic light-ion collisions.\bigskip

{\em {\it Acknowledgments. }} We thank F. Wang and S. Yang for useful discussions. This work is supported in part by the National Natural Science Foundation of China under Grants No.~12275082 and No.~12035006 (H.X.), the National Natural Science Foundation of China under 12275081 (X.W.), and the National Natural Science Foundation of China under No.\ 12135011. This work is supported in part by National Key Research and Development Program of China under Contract No.\ 2022YFA1605500, Chinese Academy of Sciences (CAS) under Grant No.\ YSBR-088.

\bibliography{Ref}

\end{document}